\documentclass{article}
\usepackage{graphicx} 
\usepackage{float}
\usepackage[a4paper, portrait, margin=1in]{geometry}
\title{The Detection of KIC 1718360, A Rotating Variable with a Possible Companion, Using Machine Learning}
\author{Jakob Roche}
\date{May 2024}

\begin{document}

\maketitle
\begin{abstract}
    This paper presents the detection of a periodic dimming event in the lightcurve of the G1.5IV-V type star KIC 1718360. This is based on visible-light observations conducted by both the TESS and Kepler space telescopes. Analysis of the data seems to point toward a high rotation rate in the star, with a rotational period of 2.938 days. The high variability seen within the star's lightcurve points toward classification as a rotating variable. The initial observation was made in Kepler Quarter 16 data using the One-Class SVM machine learning method. Subsequent observations by the TESS space telescope corroborated these findings. It appears that KIC 1718360 is a nearby rotating variable that appears in little to no major catalogs as such. A secondary, additional periodic dip is also present, indicating a possible exoplanetary companion.
\end{abstract}
\section{Introduction}
    At the time of writing, over 5,200 exoplanets have been confirmed\footnotemark, and thousands more are candidates. This amounts to a wealth of data on many different planet types, including ones not found in our solar system. However, in discovering these planets, there is still a prevalence of observing lightcurves manually. This is often a painstaking process, and one that can be automated using machine learning techniques. While the model used was originally designed to perform this task, it soon proved adept at detecting variable stars as well, which have similar lightcurve profiles to the hosts of transiting exoplanets.
    
    This paper presents the finding of a periodic dip in the lightcurve of the star KIC 1718360 that was detected in both Kepler and TESS data using a One-Class Support Vector Machine (SVM) model. This observation is indicative of rotational variability in the
    G1.5IV-V type star. The star is thought to have a size comparable to that of the sun, with an estimated size of 1.09211 \(R_\odot\)
    
\section{Procedure}
    The following procedure describes the training of the model for its original task, which was to detect exoplanetary signals in Kepler lightcurves.
    
    Given the fact that there are expected to be a high number of exoplanets in the universe\footnotemark, and the chances being relatively high that some transit signals might be overlooked if lightcurves were to be labeled as either having a transit or not\footnotemark, traditional machine learning methods could not be effectively used. These methods require a positive dataset, which in this case would be lightcurves that do contain exoplanets, and a negative dataset, which would be lightcurves that do not contain exoplanets. For the aforementioned reasons, instead of training a model on data labeled as positive or negative, alternative techniques had to be used.

    The technique used in this paper revolves around anomaly detection algorithms, namely a One-Class SVM model. In the following procedure, approximately 1,500 datapoints were provided in the form of Kepler Object of Interest (KOI) lightcurves. To begin the process of training the model, the the Pre-Search Data Conditioning Simple Aperture Photometry (PDCSAP) flux data provided by the mission was processed into a 2-dimensional array. This array contained the raw flux data for each target star in a 2-dimensional format, with each target having a separate row. This was done to put the data in a format suitable for machine learning using the selected algorithm. This data was then normalized and detrended using a Savitzky-Golay filter\footnotemark. The machine learning model used was trained to find lightcurves that have high similarity to the KOI lightcurve data. The algorithm's similarity parameters included a periodic dip in the lightcurve as well as a dip significant enough to be a planet.

    After a period of validation and parameter adjustment, the model was deployed onto the Kepler Quarter 16 data. The star KIC 1718360, among others, was flagged as being highly similar to KOI lightcurves, despite not being on the KOI catalog themselves\footnotemark. Upon manual inspection, KIC 1718360 was determined to be a probable rotating variable, since it had a signal that repeated periodically throughout the entire lightcurve.

    Upon closer inspection, the signal was found to be present in all of Kepler's observations, ranging from Kepler's first light in 2009 to the end of the Kepler mission four years later. Evidence corroborating this finding was sourced from the Transiting Exoplanet Survey Satellite (TESS). The TESS observations referenced were conducted in 2022, with virtually exactly the same period. In total, this amounts to a thirteen year period in which the same signal was detected in the lightcurve of this star. While there are no catalogued exoplanets orbiting KIC 1718360, given the lightcurve profile, it seems more probable that the main dip in the star's lightcurve is instead due to rotational variability. However, upon further inspection of the lightcurve data, there seems to be a secondary dip with a period of 1.2156 days. This could be indicative of an exoplanetary companion orbiting the star.
\section{Algorithms}
    A One-Class Support Vector Machine (SVM) was used to make predictions on the Kepler dataset. One-Class SVMs are primarily anomaly-detection algorithms, and have applications in finding outlying datapoints in much larger datasets\footnotemark. In this scenario, every lightcurve was fed into the model and treated as an individual datapoint. Outlying datapoints were treated as possible transit signals and further analyzed, while the non-outlying lightcurves were assumed to contain no transits. 

    The model used a nonlinear, polynomial kernel with a degree of 4 to run predictions. When predictions were run, the majority of lightcurves were found to be non-outliers, with a significant minority (on average approximately 6\%) of lightcurves being classified as anomalous. This method proved effective in finding periodic dips in lightcurves; a large number of catalogued variable stars were flagged by the model.
\section{Data Analysis}
    \begin{figure}[H]
        \centering
        \includegraphics[width=1\linewidth]{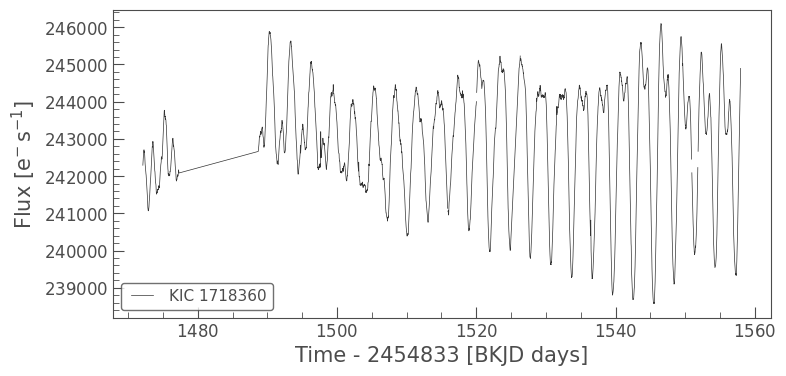}
        \caption{This is the original lightcurve detected by the model from Kepler's Quarter 16 observation. Quarter 16 lasted approximately 80 days.}
        \label{fig:Q16 Observations}
    \end{figure}
    
    Figure 1 depicts the lightcurve that the model flagged as possibly being a transit. While some long-term variation in the brightness is observed, it does not seem to impact the recurrence of a signal with a period of 2.9376 days. While the lightcurve is highly similar to rotating variable star lightcurves, this star appears in little to no major catalogs as a variable\footnotemark. While it is highly likely that it is an under-catalogued variable, the star's relative proximity to Earth (around 978 ly), and the extent to which we've catalogued nearby variables\footnotemark makes this a strange occurrence.

    \begin{figure}[H]
        \centering
        \includegraphics[width=1\linewidth]{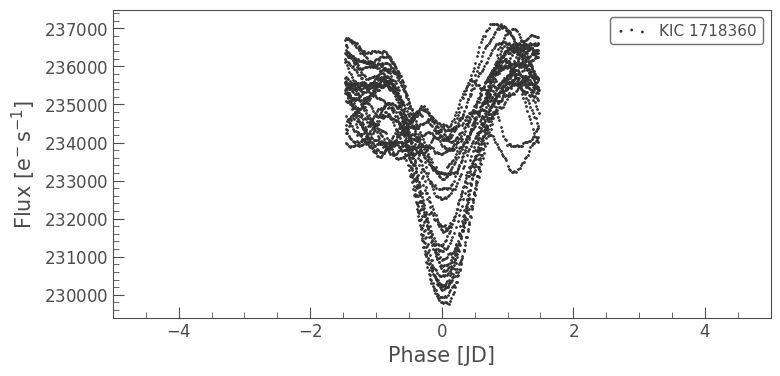}
        \caption{This is the lightcurve in Figure 1 folded and scattered by a period of approximately 2.94 days. It was folded 26 times.}
        \label{fig:Fold-Scatter Q16}
    \end{figure}
    
    Figure 2 shows the previous lightcurve folded and scattered by the suspected rotation period. This was important in confirming the periodicity of the dip. It is highly unlikely that this signal is caused by a starspot\footnotemark, as it appears not only in Kepler's 4-year observation period, but also in 2022 as well.
    
    \begin{figure}[H]
        \centering
        \includegraphics[width=1\linewidth]{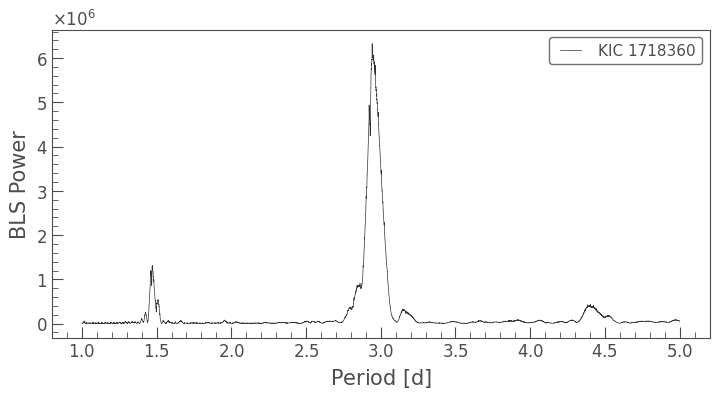}
        \caption{A Box Least Squares (BLS) periodogram created based on the Figure 1 data. Note the peak near 2.94 days, signifying a recurring signal of that period.}
        \label{fig:BLS Periodogram}
    \end{figure}
    The Box Least Squares (BLS) periodogram in Figure 3 shows a clear peak at approximately 2.9 days. This periodogram was used to pinpoint the exact rotational period of the star.
    \begin{figure}[H]
        \centering
        \includegraphics[width=1\linewidth]{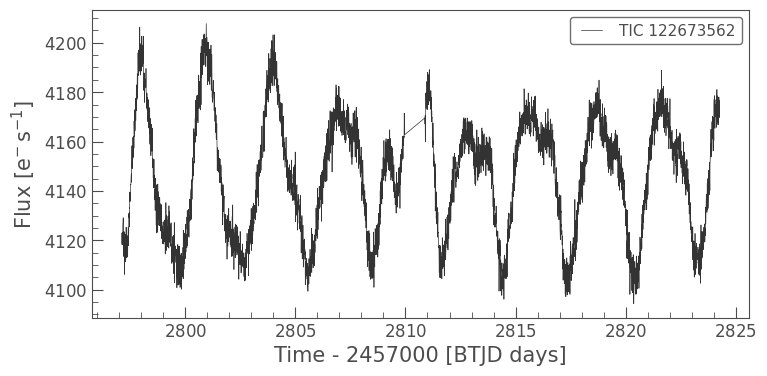}
        \caption{TESS observations of KIC 1718360. Note the same trend observed in the Kepler data is also seen here.}
        \label{fig:TESS Corroboration}
    \end{figure}

    When the TESS catalog was queried with the star's TIC identifier (TIC 122673562), the lightcurve in Figure 4 was produced. The lightcurve was produced from TESS Sector 55 observations, which took place from August 5 to September 1, 2022. The fact that exactly the same transit signal has persisted from 2009 to 2022 strongly points toward rotational variability in this star. The signal remaining unchanged for a 13 year period virtually rules out using starspots or other short-term stellar phenomena to explain the variability of this signal. Note the small dips present along the body of the lightcurve, which are especially apparent at the lightcurve's peaks. While these dips are periodic, they do not seem to be indicative of the secondary eclipse effect usually observed in eclipsing binary systems, since they do not consistently recur at the same point in relation to the other features of the lightcurve.

    \begin{figure}[H]
        \centering
        \includegraphics[width=1\linewidth]{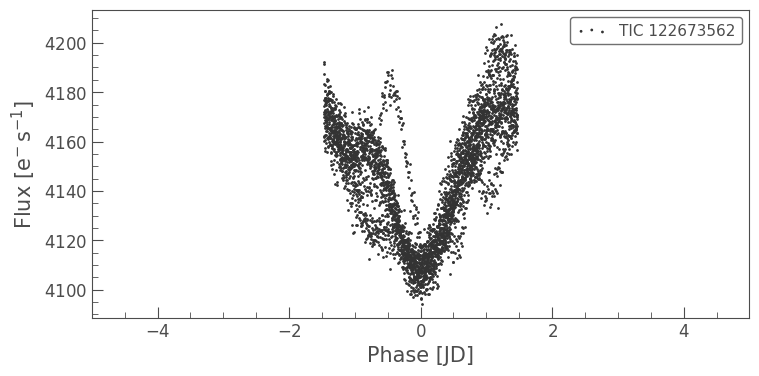}
        \caption{The lightcurve in Figure 4 folded and scattered by a period of approximately 2.94 days. Note the high similarity, yet considerably less noise, compared to the data in Figure 2.}
        \label{fig:CorroborationFoldScatter}
    \end{figure}
    \begin{figure}[H]
    \centering
    \includegraphics[width=1\linewidth]{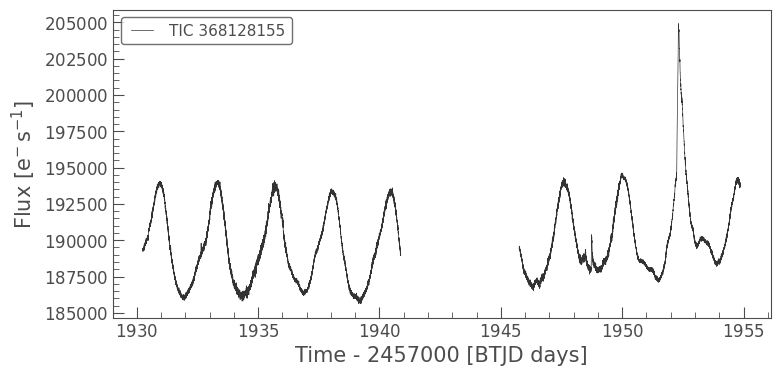}
    \caption{The lightcurve of FK Comae Berenices, a rotating variable. Note the similarity in the shape of the dip to the lightcurve in Figure 4.}
    \label{fig:enter-label}
    \end{figure}
\section{Possible Exoplanet}
    \begin{figure}[H]
        \centering
        \includegraphics[width=1\linewidth]{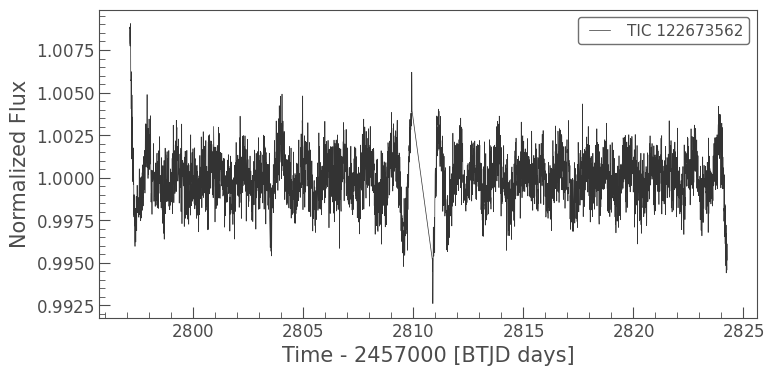}
        \caption{When the variable signal is removed from the TESS lightcurve in Figure 4, the following lightcurve is produced. Note the recurring dip with a period of 1.2156 days.}
        \label{fig:TESSFiltered}
    \end{figure}
    The lightcurve in Figure 7 is produced when the effects from the star's rotational variability are subtracted from the lightcurve in Figure 4. There is a recurring dip present in this lightcurve, with a period of approximately 1.2 days.
    \begin{figure}[H]
        \centering
        \includegraphics[width=1\linewidth]{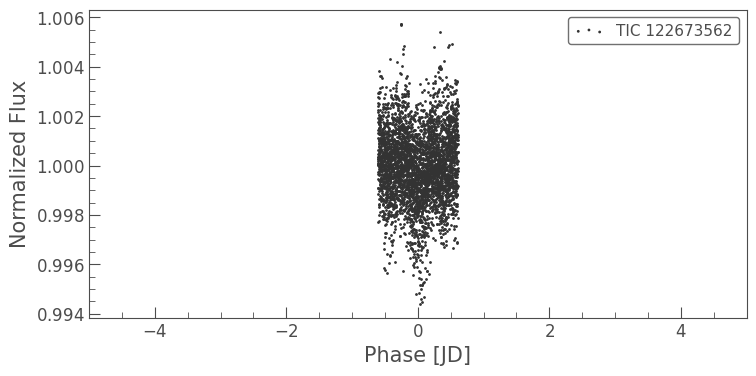}
        \caption{This lightcurve is produced when the lightcurve in Figure 7 is folded by a period of 1.2156 days.}
        \label{fig:enter-label}
    \end{figure}
    While there is a relatively high amount of noise in the lightcurve in Figure 8, there is a clear dip in the lightcurve at Phase 0. The high amount of noise is consistent with the idea that the target star exhibits stellar variability. Despite this, there appears to be a periodic dip in the data.
    \begin{figure}[H]
        \centering
        \includegraphics[width=1\linewidth]{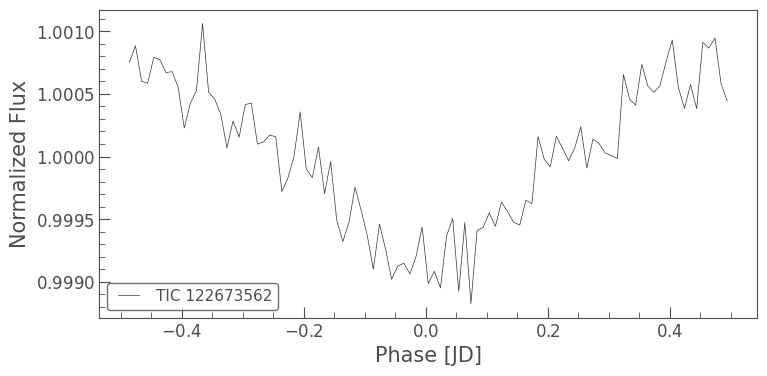}
        \caption{When the lightcurve in Figure 8 is binned to improve visual clarity, this graph is obtained.}
        \label{fig:enter-label}
    \end{figure}
    Figure 9 is obtained when the lightcurve in Figure 8 is binned. A dip can now be seen much more clearly than in Figure 8. While the dip is very slight, it still is discernible and present in all Kepler and TESS observations of the star.
\section{Parameter Analysis}
    The following are the stellar parameters calculated from the lightcurve in Figure 2.
\begin{table}[H]
    \centering
    \begin{tabular}{cccc}
         Rotation Period&  Stellar Radius&  Rotational Velocity\\
         2.9376 days&  1.09211 \(R_\odot\)&  18.81694 km/s
    \end{tabular}
    \caption{The expected stellar parameters for the suspected variable.}
    \label{tab:my_label}
\end{table}

\begin{table}[H]
    \centering
    \begin{tabular}{cccc}
         Orbital Period&  Planetary Radius&  Semi-Major Axis\\
         1.2156 days&  3.9579 \(R_\oplus\)&  0.0229 AU
    \end{tabular}
    \caption{The expected planetary parameters for possible transit. Calculated from Figure 9.}
    \label{tab:my_label}
\end{table}
    The parameters in Table 2 show that this exoplanet candidate, if it exists, must orbit in extreme proximity to its host star. However, this is not without precedent\footnotemark. Additionally, it seems to be in the expected radius range of a super-Earth.
\section{Conclusions}
    This research provides compelling evidence for the stellar variability of KIC 1718360. Analysis of the periodic dimming event in the lightcurve, observed by both the TESS and Kepler space telescopes, suggests the likelihood of a rotational period of 2.938 days for the target star.

    While it is not possible to definitively conclude the nature of the signal from lightcurve data alone, the data collected strongly suggests that the star is rotationally variable. The fact that multiple space telescopes across long periods of time corroborate the finding makes the signal unlikely to be any kind of instrumental noise or short term stellar fluctuations. Instead, it seems, there is long-term rotational variability in this star.

    Additionally, it seems there is a case to be made for the existence of a possible exoplanetary companion orbiting the star. In addition to the main dip caused by rotational variability, there seems to be a periodic secondary dip in the lightcurve of KIC 1718360, which could suggest the existence of a super-Earth orbiting this star. Further study is needed to confirm or deny the existence of this exoplanet candidate.
\section{Bibliography}

    1. Tang, Yanke, et al. "The Statistical Analysis of Exoplanet and Host Stars Based on Multi-Satellite Data Observations." Universe 10.4 (2024): 182.
    
    2. Batalha, Natalie M. "Exploring exoplanet populations with NASA’s Kepler Mission." Proceedings of the National Academy of Sciences 111.35 (2014): 12647-12654.

    3. Pearson, Kyle A., Leon Palafox, and Caitlin A. Griffith. "Searching for exoplanets using artificial intelligence." Monthly Notices of the Royal Astronomical Society 474.1 (2018): 478-491.

    4. Hippke, Michael, et al. "Wōtan: Comprehensive time-series detrending in Python." The Astronomical Journal 158.4 (2019): 143.

    5. “Kepler Object of Interest Catalog.” Caltech Archive, exoplanetarchive.ipac.caltech.edu/cgi-bin/TblView/nph-tblView?app=ExoTbls\&config=cumulative. Accessed 3 May 2024. 

    6. Erfani, Sarah M., et al. "High-dimensional and large-scale anomaly detection using a linear one-class SVM with deep learning." Pattern Recognition 58 (2016): 121-134.

    7. “AAVSO Variable Star Catalog.” Edited by Christopher Watson, The International Variable Star Index (VSX), www.aavso.org/vsx/. Accessed 3 May 2024. 

    8. Minniti, D., et al. "VISTA Variables in the Via Lactea (VVV): The public ESO near-IR variability survey of the Milky Way." New Astronomy 15.5 (2010): 433-443.

    9. Hussain, G. A. J. "Starspot lifetimes." Astronomische Nachrichten 323.3‐4 (2002): 349-356.

    10. Haghighipour, Nader. "Kepler-37b: A Moon-Sized Planet." Encyclopedia of Astrobiology. Berlin, Heidelberg: Springer Berlin Heidelberg, 2023. 1611-1612.
    
\end{document}